\documentstyle[aps,graphics,multicol,psfig,epsfig,color]{revtex}  

\begin{document}

\title{Dissecting financial markets: Sectors and states}

\author{Matteo Marsili}

\address{Abdus
Salam International Center for Theoretical Physics, Strada Costiera 11, 
34014 Trieste, Italy\\
and\\
Istituto Nazionale per la Fisica della Materia (INFM),
Unit\'a Trieste SISSA, Via Beirut 2-4, 34014 Trieste and }

\date{\today}  

\maketitle 

\begin{abstract}
By analyzing a large data set of daily returns with data clustering
technique, we identify economic sectors as clusters of assets with a
similar economic dynamics. The sector size distribution follows Zipf's
law. Secondly, we find that patterns of daily market-wide economic
activity cluster into classes that can be identified with market
states. The distribution of frequencies of market states shows
scale-free properties and the memory of the market state process
extends to long times ($\sim 50$ days). Assets in the same sector
behave similarly across states. We characterize market efficiency by
analyzing market's predictability and find that indeed the market is
close to being efficient. We find evidence of the existence of a dynamic
pattern after market's crashes.
\end{abstract}


\begin{multicols}{2}
\narrowtext           

\section{Introduction}

Thanks to the availability of massive flows of financial data,
theoretical insights on financial markets can nowadays be tested to an
unprecedented precision in socio-economic systems. This poses a
challenge which has attracted natural scientists who have pioneered an
{\em empirical} approach to financial fluctuations
\cite{Mandelbrot1,MantegnaStanley,BouchaudPotters} independent of the
econometric approach and often in contrast with the {\em axiomatic}
approach of theoretical finance \cite{Farmer,duffie}.

The empirical evidence depicts financial markets as complex
self-organizing critical systems: The statistics of real market
returns deviate considerably from the Olympic Gaussian world described
by Louis Bachelier at the turn of last century. Rather Mandelbrot
\cite{Mandelbrot2} observed that fractal (Levy) statistics gives a
closer approximation, even though that is not a satisfactory
model\cite{Mandelbrot1,MantegnaStanley}. Market returns display
scaling\cite{MantegnaStanley}, long range volatility correlations
and evidence of multiscaling \cite{multisc} have also
been discussed. Such features evoke the theory of critical phenomena
in physics, which explains how quite similar features may emerge from
the interaction of many microscopic degrees of freedom and statistical
laws. Indeed financial markets {\em are} systems of many interacting
degrees of freedom (the traders) and there are very good theoretical
reasons to expect that they operate rather close to criticality
\cite{MEM}. These expectations have been substantiated by microscopic
agent based market models\cite{CCMZ,CMZ01,BMRZ}: The picture offered
by these {\em synthetic markets} is one where speculation drives
market to information efficiency -- i.e. to a point where market
returns are unpredictable. But the point where markets become exactly
efficient is the locus of a {\em phase transition}. Close to the phase
transition the behavior of synthetic markets is characterized by the
observed stylized facts -- fat tails and long range correlations --
whereas far from the critical region the market is well described in
terms of random walks (see Ref. \cite{CCMZ} for a non technical
discussion).

Work has however been mostly confined on single assets or
indices. Recently ensembles of assets and their correlations have
become the focus of quite intense interest. On one side the role of
random matrix theory has been realized as a tool for understanding how
noise dresses financial correlations \cite{Focus} how one can undress
them \cite{GM}, how clustering techniques can help understanding the
structure of correlation \cite{Mantegna}, and the impact of such
consideration on portfolio optimization \cite{Gopiport}.

Here we report findings that strongly support the view of a
self-organized critical market. We show that long range correlations
and scale invariance extends both across assets and, in the behavior
of the ensemble of assets, across frequencies. More precisely, we
apply a novel parameter free data clustering method \cite{GM,mldc} to a
large financial data set \cite{data_set} in order to uncover the
internal structure of correlations both across different assets and
across different days. We identify statistically significant
classifications of assets in correlated {\em sectors} and of daily
profiles of market-wide activity in market {\em states}. Both the
statistics of sector sizes and of state sizes shows scale free
properties.

Determining market's states is an important achievement both
theoretically and practically: The concept of a state which codifies
all relevant economic informations is the basis of many theoretical
models of financial markets. But practically every day traders
experience a quite different reality: The market place is
flooded with massive flows of information of which it may be hard to
say what is relevant and what is irrelevant. It is by no means obvious
that something like market states exists at all and even if they exist
the problem becomes that of identifying them. Our aim is to give a
practical answer to these questions. We shall keep our discussion as
simple as possible, relegating technical details in notes and in the
appendix.

\section{The method and the data set}

The data clustering method that we use has been recently proposed in
Ref. \cite{GM}. In brief, it is based on the simple statistical
hypothesis that {\em similar objects have something in common}. 
It is possible to compute the likelihood that a given data set
satisfies this hypothesis and hence to look for the most likely 
cluster structure. A precise definition is given in the appendix and
for more details we refer the interested reader to
Refs. \cite{GM,mldc}.  Let us only mention that this method overcomes
several limitation of traditional data clustering approaches, such as the
needs of pre-defining a metric, fixing {\em a priori} the number of
clusters or tuning the value of other parameters\cite{mldc}. 

The data set covers a period from 1st January 1990 to 30th of April
1999 and it reports daily prices (open, hi, low, close) for $7679$
assets traded in the New York Stock Exchange \cite{data_set}. 
The number of assets actually traded varies with time. Hence we mainly
focus on a subset of the $2000$ most actively traded assets (see
{\tt http://www.sissa.it/dataclustering/fin/} for the detailed list 
of assets considered, as well as for further informations).

Our goal is to investigate the {\em internal} structure of
correlations hence we first normalize the raw data \cite{norma} in
order to eliminate common trends and patterns both across assets and
across different days.  
This procedure eliminates for example the so-called ``market mode'',
i.e. the constant correlation of individual asset's returns with the
so-called ``market's return''.

\section{Market sectors: Scale free market structure}

\begin{figure}
\centerline{\psfig{figure=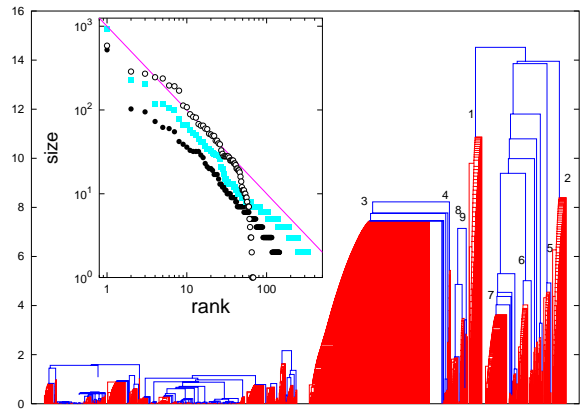,width=8cm}}
\caption{Dendrogram of the cluster structure of correlated sectors
resulting from hierarchical clustering algorithm. Assets are reported
along the horizontal axis and red shapes correspond to clusters of
correlated assets. The height of a shape is the contribution to the
log-likelihood of the corresponding cluster of assets. See the
appendix for more details. The cluster structure
is statistically significant because the noise level corresponding to
uncorrelated data would show structures with a log-likelihood of at
most $0.1$, three orders of magnitude smaller.  The classification in
sectors has a large overlap with economic sectors. For example,
clusters 1 and 2 contain firms in the electric sector and computers
respectively. Cluster 4 is the sector of gold, 5 is composed of banks,
8 contains oil and gas firms, 9 petroleum. Clusters 3, 6 and 7 are
mixed clusters (more details are available at 
{\tt http://www.sissa.it/dataclustering/fin/}). Inset: Distribution of
correlated sector sizes for $2000$ ($\bullet$) and $4000$ ($\Box$)
assets. The distribution of the size of economic sectors ($\circ$), as
defined by the (first two digits of the) SIC codes, for the same
$4000$ assets is shown for comparison. The line (drawn as a guide to
the eyes) has slope $-1$.}
\label{figassets}
\end{figure}

We first apply data clustering to group assets with a similar economic
dynamics in sectors of {\em correlated} assets (see appendix). This
classification reveals a rich structure. The clusters giving the
largest contributions to the log-likelihood clearly emerge from the
noisy background in Fig. \ref{figassets}. We find a large overlap with
the sectors of economic activity defined by the Standard Industrial
Classification (SIC) codes (see caption of Fig. \ref{figassets}). But
we also find significant correlations between assets with widely
different SIC. This has practical relevance for risk management of
large portfolios which cannot be handled all at once. Indeed rather
than splitting the problem according to economic sectors (defined by
the SIC) it is preferable to use our classification in correlated
sectors. The difference of the two classifications is also revealed by
a Zipf's plot of the size of sector against its rank (see inset of
Fig. \ref{figassets}). The distribution of correlated sector sizes
follows Zipf's law to a high accuracy, i.e. the number ${\cal N}(n)$
of sectors with more than $n$ firms (i.e. of size larger than $n$) is
inversely proportional to $n$. Note that the scale free distribution
of sector sizes is not due to an analogous property of {\em
fundamentals}. Indeed the rank plot of economic sector sizes bends in
log-log scale. This suggests that Zipf's law arises as a dynamical
consequence of market interaction.

The scale invariant behavior is robust with respect to the subset of
assets taken: The same behavior is found considering the
$1000,~2000$ or $4000$ most actively traded assets, in that
period or $443$ assets in the S\&P500 index (see
Ref. \cite{GM}). In addition we find, as in Ref. \cite{GM}, that the
correlation $c_s$ inside sector $s$ (see appendix) scales with its
size $n_s$ with a law $c_s\sim n_s^\gamma$ with $\gamma\simeq 1.66$.

We finally remark that this property is not an artifact of the
method. Indeed the distribution of eigenvalues of the correlation
matrix shows a similar broad distribution, even though that is
affected by considerable noise dressing \cite{Focus}. A factor model
which takes into account a large enough number of principal components
(corresponding to the largest eigenvalues) reproduces the same
features\footnote{In our case $\approx 30$ eigenvalues of the
correlation matrix are significantly outside the noise band predicted
by Random Matrix Theory \cite{Focus}. With a correlation matrix which
retains the structure of the first $\sim 20$ principal components
(considering the remaining components as uncorrelated noise) we found
a quite similar cluster structure.}.

\section{Market states}

\begin{figure}
\centerline{\psfig{figure=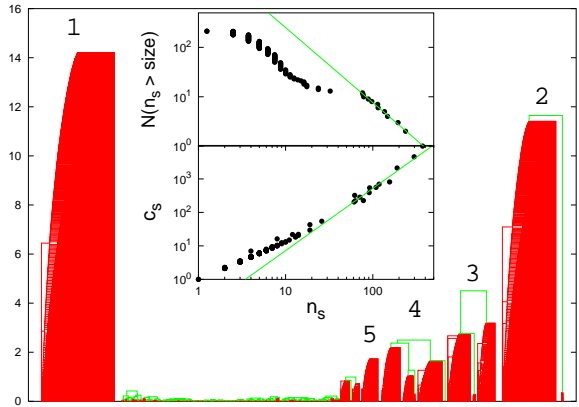,width=8cm}}
\caption{Same plot as Fig. 1 for days: Clusters of days
identify market states. We identify states (see labels) as groups of
correlated clusters of days.  Inset: Distribution of cluster sizes,
i.e. of the frequency with which states occur (top) and correlation
$c_s$ inside each cluster (bottom).}
\label{figdays}
\end{figure}

Are there well defined patterns of daily market-wide economic
performance? In order to answer this question, rather than classifying
assets according to their temporal evolution, we can classify days
according to the performance of different assets. Fig. \ref{figdays}
implies that, above a noisy background, a meaningful classification of
the daily profiles of market activity exists. Clusters of days can be
identified with different patterns of market wide activity -- or
market states. Quite remarkably, the maximum likelihood classification
in market states shows scale free features, for large clusters
(frequent patterns of market activity). The number of patterns which
occur more than $d$ days behaves as ${\cal N}(d)\sim d^{-1.5}$ for the
most frequent patterns (inset top). There is a clear crossover in the
plot of cluster's correlation versus cluster size which distinguishes
the meaningful clusters (patterns) from a random noise background
(inset bottom).

From a sample of $2000$ assets over $T=2358$ days we identify $5$
different states -- characterized by similar profiles of market
activity -- plus a sixth random state (see 
Fig. \ref{figdays}). We assign an integer $\omega(t)$ between $1$ and
$6$ to each day $t$, which is the state which occurred in that day.

\begin{figure}
\centerline{\psfig{figure=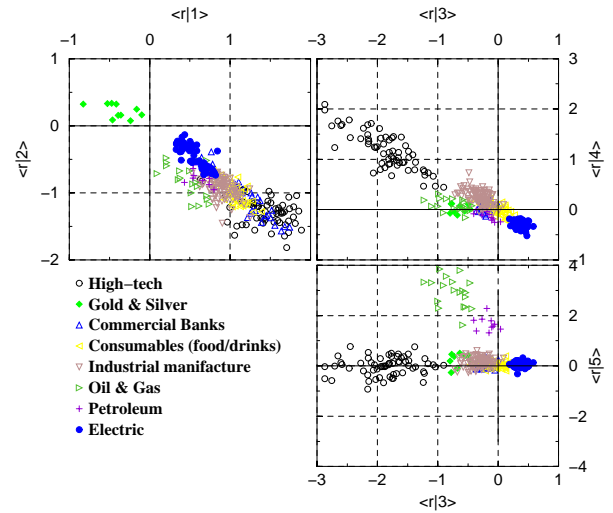,width=8cm}}
\caption{Performance of the market in different states. Each asset $i$
corresponds to a point whose coordinates are the average returns
$(\langle{r_i|\omega}\rangle,\langle{r_i|\omega'}\rangle)$ of asset
$i$ in states $\omega$ and $\omega'$. Assets in different sectors are
plotted differently.}
\label{figcross}
\end{figure}

We are then in a position to analyze market performance in different
states. Fig. \ref{figcross} shows the (non normalized) average daily
returns of different asset in different states. We find that market's
behavior in states 1 and 2 are anti-correlated: Those assets which go
up in state 1 go down in state 2, on average. Fig. \ref{figcross} also
shows that assets in the same sector as defined above have a similar
behavior. So, for example, while most of the assets go up in state 1
and down in state 2, the cluster of assets of Gold and Silver mining
has an opposite behavior. State 3 is clearly characterized by a fall
of High-tech companies and a mild rise in the electric sector. An
opposite behavior takes place in state 4, whereas state 5 is dominated
by the a marked rise of Oil \& Gas, and Petroleum refining companies
\cite{data_set}. 

These results are remarkably stable with respect to the definition of
the time window where the analysis is performed \cite{stability}.

\subsection{Predictability and market efficiency}

Clustering the market's dynamics leaves us with the sequence
$\omega(t)$ of the states of the market in different days
$t=1,\ldots,T$. This allows us to pose interesting questions on
predictability and market's information efficiency.  

Let us first ask: Is it possible to predict the
state $\omega'$ of the market tomorrow, given the state $\omega$ of
the market today? In order to answer this question we estimate the
probability
\[
P_1(\omega'|\omega)=\sum_{t=1}^{T-1} 
\delta_{\omega(t),\omega}\delta_{\omega(t+1),\omega'}
/\sum_{t=1}^{T-1} 
\delta_{\omega(t),\omega}
\]
of transition from state $\omega$ to state $\omega'$. It turns out
that both the classification in states and the transition matrix
$P_1(\omega'|\omega)$ are very stable with respect to the definition of
the time window \cite{stability}. This means that they both vary very
slowly in time. Hence we shall neglect their variation in time
henceforth.

{\em If} the process $\omega(t)$ were Markovian, its predictability
could be quantified by the characteristic time $\tau$ of convergence
to the stationary state. This is related to the second largest
(in absolute value) eigenvalue $\lambda$ of the matrix
$P_1(\omega'|\omega)$ by $\tau=-1/\log|\lambda|$. We find $\tau\approx
0.54$ days -- a value which would occur by chance, if there were no
correlations, in one out of $10^7$ cases\footnote{This conclusion was
reached considering the characteristic times $\tau$ for symbolic
sequences $\tilde\omega(t)$ generated by randomly reshuffling
days. These times are distributed around $\tau\approx
0.33$ with a spread $\delta\tau\approx 0.04$. The analysis of the tail
of the distribution allows to estimate the likelihood of $\tau\simeq
0.54$ for the real sequence.}.  Statistical prediction is possible.

\begin{figure}
\centerline{\psfig{figure=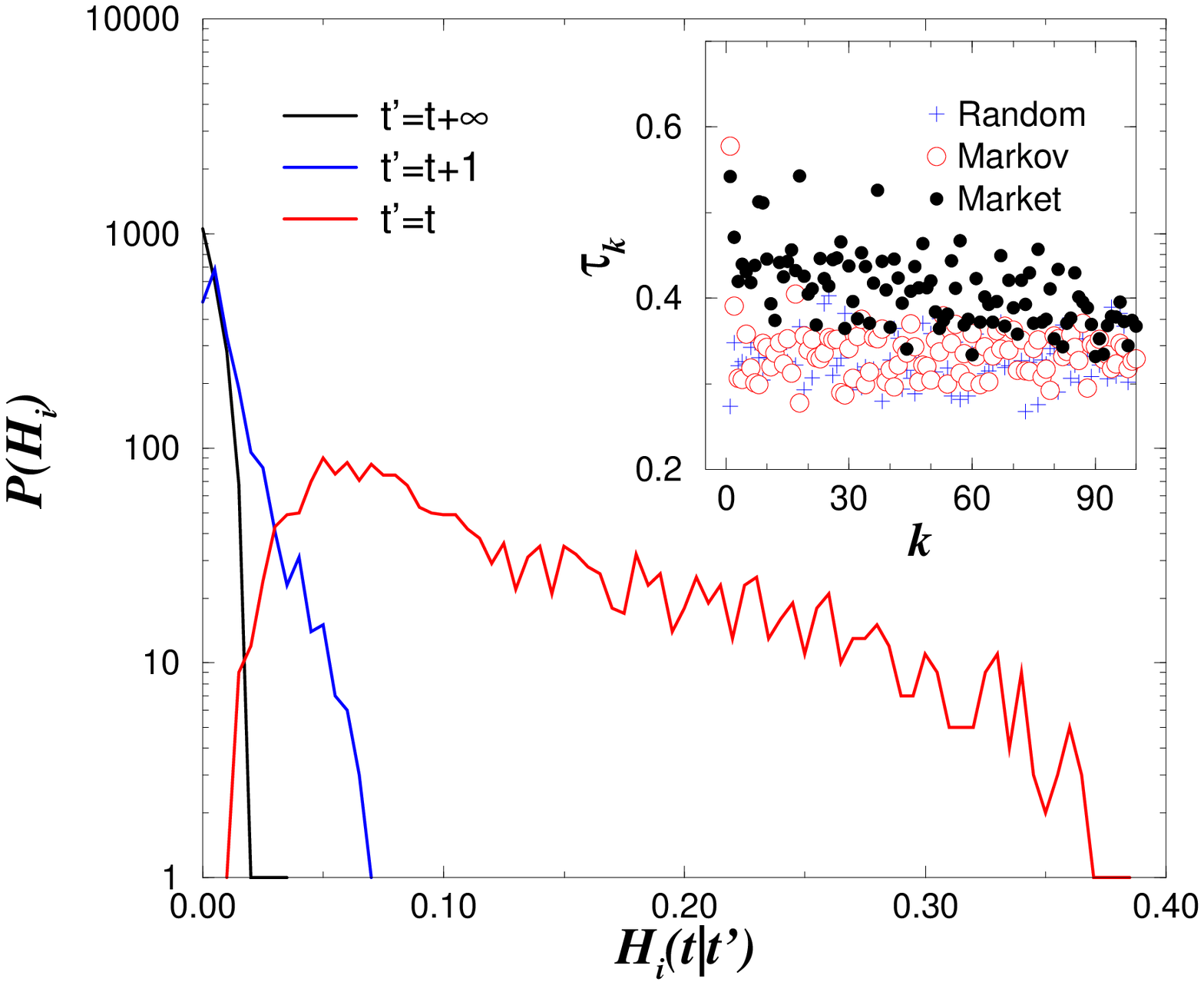,width=8cm}}
\caption{Distribution of predictability $H_i(t'|t)$ for $t'=t,~t+1$
and $t+\infty$. The noise background predictability $H_i(t+\infty|t)$
is estimated drawing $\omega(t +\infty)$ at random from the
populations of states.  Inset: Characteristic times $\tau_k$ for
transitions over $k$ days for the real sequence $\omega(t)$
($\bullet$), a random sequence ($+$) and a Markov chain sequence
($\circ$) generated with the transition probability
$P_1(\omega'|\omega)$ estimated from $\omega(t)$. The random sequence
($+$) represents the noise background. For a Markov chain $\tau_k$
($\circ$) is significantly above the noise level only for $k=1$. 
For the real market process $\tau_k$ is well above the noise level
up to $k\approx 50$.}
\label{figPH}
\end{figure}

Can we predict market's returns on the basis of these results?
Fig. \ref{figcross} shows that average returns $\langle
r_i(t)|\omega(t)\rangle$ conditional on the state $\omega(t)$ of the
market contain non-trivial information. However this information is
not available for trading in day $t$. But if we know the transition
matrix $P_1(\omega'|\omega)$ we can estimate the expected return of
asset $i$ tomorrow given the state $\omega$ today:
\[
\langle r_i(t+1)|\omega(t)\rangle=\sum_{\omega'}
\langle r_i(t+1)|\omega(t+1)=\omega'\rangle P_1(\omega'|\omega(t)).
\]
A natural measure of predictability, inspired by works on theoretical
models \cite{CM,CMZ,CCMZ,BMRZ}, is the averaged signal-to-noise ratio
defined as:
\[
H_i(t'|t)= 
\sqrt{\sum_\omega \rho_\omega
\frac{\langle \delta r_i(t')|\omega(t)=\omega\rangle^2}
{\langle\delta r_i^2|\omega\rangle}}
\]
where $\delta r_i(t)=r_i(t)-\langle r_i\rangle$ and $\rho_\omega$ is
the frequency with which state $\omega$ occurs.  The distribution of
$H_i$ across assets is shown in Fig. \ref{figPH} for $t'=t$, $t'=t+1$
and $t'=t+\infty$.  The latter gives a benchmark of the
background noise level. We find $H_i(t|t)\gg H_i(t+\infty|t)$ for
several assets $i$: the knowledge of $\omega(t)$ {\em before} day $t$
provides significant predictive power on excess returns. That same
information is much less useful the day after, since $H(t+1|t)$ is
only slightly above the noise level. This is a further indication that
the financial market is close to information efficiency, but not quite
unpredictable. In reality the transition matrix $P_1(\omega'|\omega)$
changes slowly in time. Hence this conclusion provides an ``upper
bound'' for the market's predictability (when measured out-of-sample):
Real markets are therefore even closer to efficiency.

If $\omega(t)$ were a Markov process, the characteristic time $\tau_k$
for transitions $\omega(t)\to\omega(t+k)$ over $k$
days\footnote{$\tau_k$ is computed in the same way as $\tau=\tau_1$
above, from the matrix $P_k(\omega'|\omega)$ of transition
probabilities $\omega(t)=\omega\to\omega(t+k)=\omega'$ in $k$ days. 
For a Markov process this matrix is the $k^{\rm th}$ power of the
matrix $P_1(\omega'|\omega)$ and its eigenvalues are given by
$\lambda_k=\lambda_1^k$.}
should decrease with $k$ as $\tau_k=\tau_1/k$. A prediction of the
future state of the market, which is significantly better than a
random draw, would only be possible on a time horizon of one day, if
the process were Markovian. The inset of Fig. \ref{figPH} shows that
$\tau_k$ remains significantly above the noise level almost up to
$k\approx 100$ days!  This means that $\omega(t)$ carries significant
information about the future state $\omega(t+k)$ of the market, even
after $k\approx 50$ days. The slow decay of $\tau_k$ is a further
signature of the presence of long range correlations.

During the period we have studied, two major extreme events occurs:
the 27 October 1997 and the 31 August 1998 crashes.  The state process
$\omega(t)$ is different before the crash, but is quite similar after
it. The strings of states, starting from the day of the crash, read
$2136613611\ldots$ and $2126614633\ldots$ in the two cases. This is a
significant similarity\footnote{Only two other string of the type
$21x661$ occurred in the process but the starting days were Fridays
(90/04/27 and 90/05/25) and not Mondays. Note furthermore that
normalization \cite{norma} removes the collective component of the
dynamics and it ensures that crash days appear with the same weight as
normal days in the analysis.}. This suggests the existence of a
particular dynamical pattern with which markets respond to extreme
events (see also Ref. \cite{Omori} on this).

\section{Conclusion and outlook}

In conclusion we show that both the {\em horizontal} clustering of
assets in correlated sectors and the {\em vertical} classification of 
market-wide economic performance in market states, reveal a scale free
structure (see Figs. \ref{figassets}, \ref{figdays}). The emergent
picture poses quite severe constraints on multi-asset agent based
modeling, which we believe will disclose important information on how
real markets work. This expectation is based on the fact that
scale-free statistical behavior is a signature of interaction
mechanisms which is rather insensitive to microscopic details.

Furthermore, the identification of market states allows us to
precisely quantify informational efficiency by computing the market's
predictability, thereby establishing a direct contact between the
empirical world and the realm of theoretical models. In particular we
find that, as expected, markets are close to information efficiency.

We find that correlated sectors have a large overlap with sectors of
economic activity. In the same way, it would be interesting to
understand how states are correlated with economic information and the
news arrival process. 

In a wider context, we have discussed an unsupervised approach to the
study of a complex system. Be it a stock market, the world economy,
urban traffic network, a cell of a living organism or the immune
system, the complex system can be considered as a black box.  We show
how a series of simultaneous measures in many different ``points''
of the system allows one to identify its {\em parts} and its {\em
states}.

A black box approach to a financial market or to a cell, which
neglects all of economics and finance or of biology and genetics and
relies only on empirical data, may lead to misleading results
specially if the data set is incomplete. Still, we believe, it has the
potential of uncovering collective aspects which can hardly be derived
in a theoretical bottom-up approach.

\appendix
\section{Maximum likelihood data clustering}
\label{mldc}

Consider a set of $N$ objects each of which is defined in terms of $D$
measurable features, so that each object is represented by a vector
$\vec \xi_i\in R^D$, $i=1,\ldots,N$. We assume for simplicity that data
are normalized: $\vec \xi_i\cdot \vec e=0$ where $\vec e=(1,1,\ldots,1)$
and $\|\xi_i\|^2=\vec \xi_i\vec \xi_i=1$.

In our case, when identifying sectors, the objects are assets and
$N=A$, the number of assets. Their features are the daily returns in
each day $t$ and $D=T$. The $t^{\rm th}$ component of $\vec \xi_i$ is
$x_i(t)/\sqrt{T}$. When identifying states instead objects are days
and features are assets (i.e. $N=T$ and $D=A$). The $i^{\rm th}$
component of $\vec \xi_t$ is $x_i(t)/\sqrt{A}$. 

The problem of classifying  $N$ objects into different classes
goes under the name of data clustering.  Naively one would like to
have similar objects classified in the same cluster, but in practice
one faces a number of problems: What does it mean similar?  What is
the ``right'' number of clusters?  Which principle to follow?  We
resort to a recent data clustering technique \cite{GM,mldc} based on
the maximum likelihood principle and a simple statistical hypothesis:
{\em similar objects have something in common}. In mathematical terms,
we let $s_i$ be the label of the cluster to which object $i$ belongs,
and $A_s=\{i:~s_i=s\}$ be the set of objects with $s_i=s$. We assume
that
\begin{equation}
\vec \xi_i = g_{s_i}\vec\eta_{s_i}+\sqrt{1-g_{s_i}^2}\vec\epsilon_i.
\label{ansatz}
\end{equation}
Here $\vec \eta_s$ denoted the {\em common} component shared by all
objects $i\in A_s$ and $g_s\ge 0$ weights the common component against
the individual one $\vec\epsilon_i$. Eq. (\ref{ansatz}) is the 
statistical hypothesis where $g_s$ and $s_i$ are the parameters to be
fitted. Assuming further that both $\vec \eta_s$ and $\vec \epsilon_i$
are Gaussian vectors in $R^D$, with zero average and unit variance
($E[\|\eta_s\|^2]=E[\|\epsilon_i\|^2]=1$) makes it possible to compute
the likelihood of the parameters ${\cal G}=\{g_s\}$ and ${\cal
S}=\{s_i\}$ (see Ref. \cite{GM} for details). The likelihood is
maximal when
\begin{equation}
g_s=\sqrt{\max\left[0,\frac{c_s-n_s}{n_s^2-n_s}\right]}
\end{equation}
where $n_s=|A_s|$ is the number of objects in
cluster $s$ and
\[
c_s=\sum_{i,j\in A_s} \vec \xi_i\vec \xi_j
\]
is the total correlation inside cluster $s$.
The maximum log-likelihood per feature takes the form
\[
{\cal L}_c({\cal S})=\frac{1}{2}\sum_{s:~n_s>1}\max
\left[0,\log
\frac{n_s}{c_s}+(n_s-1)\log\frac{n_s^2-n_s}
{n_s^2-c_s}\right].
\]
Note that a cluster with a single isolated object ($n_s=c_s=1$), or a
cluster of uncorrelated objects ($c_s=n_s$) gives a vanishing
contribution to the log-likelihood. 

Several algorithms for finding an approximate maximum of ${\cal L}_c$
over the space of cluster structures ${\cal S}$ have been discussed in
Ref. \cite{mldc}. We used both hierarchical clustering and simulated
annealing algorithms, which yield quite similar results (the codes are
available on the Internet \cite{data_set}).

Figures \ref{figassets} and \ref{figdays} are a graphic representation
of the hierarchical clustering algorithm: It starts from $N$ clusters
composed of a single object and it produces a sequence of cluster
structures. At each iteration, two clusters of the configurations with
$K$ clusters are merged so that the log-likelihood of the resulting
configuration with $K-1$ clusters is maximal. This procedure starts
with $K=N$ and it stops with $K=1$, when a single cluster is
formed. The log-likelihood of the cluster structure is ${\cal L}_c=0$
when $K=N$, it decreases with $K$ and it reaches a minimum for an
intermediate value of $K$. Then it increases again and reaches ${\cal
L}_c=0$ when $K=1$, because of data normalization. 

The graphs report the log-likelihood of each cluster on the $y$ axis.
The initial configuration corresponds to $N$ points aligned on the $x$
axis (zero log-likelihood). Each merge operation is represented
graphically by a link between the merging clusters and the new
cluster. Hence as the log-likelihood decreases structures above the
$x$ axis start to form. Red links are merging steps which increase the
log-likelihood. Blue links corresponds to situation where the
log-likelihood of the union of the clusters is larger than that of
each part but it is smaller than their sum (hence the total
log-likelihood decreases). Hence statistically relevant clusters
appear as the large red structures in the plot.

\end{multicols}  


\begin{thebibliography}{99}

\bibitem{Mandelbrot1} Mandelbrot, B. B., {\em Fractals and
Scaling in Finance}, Springer-Verlag (New York 1997).

\bibitem{MantegnaStanley} R.N. Mantegna and H.E. Stanley, {\em Introduction
to Econophysics: Correlations and Complexity in Finance}, Cambridge
Univ. Press (Cambridge UK, 1999). 

\bibitem{BouchaudPotters} J.-P. Bouchaud and M. Potters, {\em Theory of
Financial Risk: From Statistical Physics to Risk Management},
Cambridge Univ. Press (Cambridge UK, 2000)

\bibitem{Farmer} J.D. Farmer, Physicists Attempt to Scale the Ivory
Towers of Finance , Computing in Science and Engineering (IEEE),
{\bf 1} 1999, 26-39.

\bibitem{duffie} J.Y. Campbell, A.W. Lo, and A.C. MacKinlay, {\em The
Econometrics of Financial Markets}, Princeton Univ. Press (Princeton
N.J., 1997).

\bibitem{Mandelbrot2} B.B. Mandelbrot, The Variation of Certain
Speculative Prices, J. Business, Vol. 36, 1963, pp. 394 419.

\bibitem{multisc} S. Ghashghaie et al., {\em Turbulent Cascades in Foreign
Exchange Markets}, Nature, {\bf 381}, 767 (1996).

\bibitem{MEM} D. Challet, M. Marsili and Y.-C. Zhang, {\em Modeling
market mechanism with minority game}, Physica A {\bf 276}, 284
(2000).

\bibitem{CCMZ} D. Challet et al., {\em From Minority Games to real
markets}, Quantitative Finance {\bf 1}, 168 (2001).

\bibitem{CMZ01} D. Challet, M. Marsili and Y.-C. Zhang, {\em Stylized
facts of financial markets and market crashes in Minority Games},
Physica A {\bf 294}, 514 (2001).

\bibitem{BMRZ} J. Berg et al. {\em Statistical mechanics of asset
markets with private information}, Quantitative Finance {\bf 1}, 203
(2001).

\bibitem{Focus} Laloux et al., {\em Noise Dressing of Financial
Correlation Matrices}, Phys. Rev. Lett. {\bf 83}, 1467 (1999);
V. Plerou et al. {\em Universal and Nonuniversal Properties
            of Cross Correlations in Financial Time
            Series}, {\em ibid} 1471.

\bibitem{GM} L. Giada, M. Marsili, {\em Data clustering and noise
undressing of correlation matrices}, Phys. Rev. E {\bf 63}, 1101
(2001).

\bibitem{Mantegna} R.N. Mantegna, {\em Hierarchical structure in
financial markets}, Eur. Phys. J. B {\bf 11} , 193 (1999).

\bibitem{Gopiport} P. Gopikrishnan et al., {\em Quantifying and
interpreting collective behavior in financial markets}, Phys. Rev. E
{\bf 64}, 035106 (2001).

\bibitem{mldc} L. Giada, M. Marsili, {\em Algorithms of maximum
likelihood data clustering with applications}, eprint cond-mat/0204008
(2002).

\bibitem{data_set} The data set was made available by courtesy of
R. N. Mantegna. The tic symbols of the subset of assets considered,
the detailed cluster structures of sectors and states and other
informations are available at {\tt
http://www.sissa.it/dataclustering/fin/}.

\bibitem{norma} Let $x_i^{(0)}(t)=\log p_i^{\rm open}(t)/p_i^{\rm
close}(t)$ be the return of asset $i=1,\ldots, A$ in day
$t=1,\ldots,T$. We set 
\begin{eqnarray*}
x_i^{(2k+1)}&=&\frac{x_i^{(2k)}(t)-\langle x_i^{(2k)}\rangle}
{\sqrt{\langle(x_i^{(2k)}-\langle x_i^{(2k)}\rangle)^2\rangle}}\\
x_i^{(2k+2)}&=&\frac{x_i^{(2k+1)}(t)-\overline{x_i^{(2k)}}}
{\sqrt{\overline{(x_i^{(2k)}-\overline{x_i^{(2k)}})^2}}}
\end{eqnarray*} 
where $\langle\ldots\rangle=\sum_{t=1}^T(\ldots)/T$ is
a time average and
$\overline{(\ldots)}=\sum_{i=1}^A(\ldots)/A$ denotes the
average over assets. As in M. B. Eisen et al., 
[Proc. Natl. Acad. Sci. USA, {\bf 95}, 14863 (1998).], the normalized
data $x_i(t)$, is obtained as the limit of $x_i^{(n)}(t)$ as
$n\to\infty$. In practice the iteration was stopped after a given
accuracy was reached. This procedure does not affect significantly the
results. Indeed the first step of normalization eliminates most of the
global patterns. For missing values we assumed $x_i(t)=0$ if asset $i$
were not traded on day $t$.

\bibitem{stability} In order to asses the stability of the results we
repeated the classification of days for the first (from Jan. '90 to
Aug. '94) and the second (Sep. '94 to Apr. '99) halves of the time
series. We found dendrograms quite similar to those in
Fig. \ref{figdays} with two main dominant states. Clustering again
days into $6$ states, we found two new sequences
$\omega_{<}(t)$ for $t=1,\ldots,T/2$ and $\omega_{<}(t)$ for
$t=T/2+1,\ldots,T$. We found that $\omega_{<}(t)=\omega(t)$ in $73$\%
of cases and $\omega_{>}(t)=\omega(t)$ in $82$\% of cases, where
$\omega(t)$ is the state occurring in day $t$ according to the
analysis of the whole time series.

\bibitem{CM} D. Challet, M. Marsili, {\em Phase transition and symmetry
breaking in the minority game} Phys. Rev. {\bf E 60}, R6271 (1999).

\bibitem{CMZ} D. Challet, M. Marsili, R. Zecchina, {\em Statistical
mechanics of systems with heterogeneous agents: Minority games},
Phys. Rev. Lett. {\bf 84}, 1824 (2000).

\bibitem{Omori} F. Lillo, R.N. Mantegna, {\em Omori law after a
financial market crash}, e-print cond-mat/0111257 (to appear in Physica A).

\end{thebibliography}
\end{document}